\documentclass[PRD,showpacs,preprint,floatfix]{revtex4} 

\usepackage{hyperref}
\usepackage{epsfig} 
\usepackage{graphicx}
\usepackage{amsfonts}
\usepackage{amssymb}
\usepackage{amsbsy}
\usepackage{amsmath}
\usepackage{latexsym}

\newcommand{\ba}{\begin{array}}
\newcommand{\ea}{\end{array}}

\newcommand{\bx}{\nabla^2}
\newcommand{\tR}{{\tilde{R}}}

\newcommand{\be}{\begin{equation}}
\newcommand{\ee}{\end{equation}}
\newcommand{\bea}{\begin{eqnarray}}
\newcommand{\eea}{\end{eqnarray}}
\newcommand{\beal}{\setcounter{letter}{1} \begin{eqnarray}}
\newcommand{\eeal}{\addtocounter{equation}{1} \end{eqnarray}}
\newcommand{\none}{\nonumber \\}
\newcommand{\req}[1]{Eq.(\ref{#1})}

\newcommand{\larrow}{\,\,\,\,\hbox to 30pt{\rightarrowfill}
\,\,\,\,}
\newcommand{\slarrow}{\,\,\,\hbox to 20pt{\rightarrowfill}
\,\,\,}

\newcommand{\half}{{\frac{1}{2}}}

\renewcommand*{\c}{\text{c}}

\def\be {\begin{equation}}
\def\ee  {\end{equation}}
\def\bea {\begin{eqnarray}}

\def\eea {\end{eqnarray}}

\renewcommand*{\c}{\text{c}}

\def\be {\begin{equation}}
\def\ee  {\end{equation}}
\def\bea {\begin{eqnarray}}
\def\eea {\end{eqnarray}}

\begin{document}
\title{Quantum Mechanics of the Interior of Radiating 2-D Black Holes}
\author{Jack Gegenberg }
\email{geg@unb.ca}
\affiliation{ Department of Mathematics and
Statistics, University of New Brunswick, Fredericton, NB, Canada E3B 5A3} 
\author{Gabor Kunstatter}
\email{g.kunstatter@uwinnipeg.ca}
\affiliation{ Department of Physics, University of Winnipeg and Winnipeg Institute for Theoretical Physics, Winnipeg, Manitoba,Canada R3B 2E9}
\author{Tim Taves}
\email{timtaves@gmail.com}
\affiliation{ Department of Physics, University of Manitoba and Winnipeg Institute for Theoretical Physics, Winnipeg, Manitoba,Canada R3T 2N2 }
\date{\today}
\begin{abstract}
We study the homogeneous sector of the RST model describing the gravitational dynamics, including back-reaction, of radiating 2-d black holes. We find the exact solutions both in conformal gauge and in time-parametrized form, isolate the black hole sector of the classical phase space and quantize the near singularity dynamics in conformal gauge. We show that different choices of measure and different self-adjoint extensions can lead to inequivalent quantum theories, all of which resolve the singularity. For a specific range of extension parameters, the Hamiltonian spectrum admits bound states that correspond physically to stable remnants. Finally, we argue that our work provides a good starting point for quantization of the full homogeneous theory using both reduced and  Dirac quantization.
\end{abstract}

\maketitle
\tableofcontents
\section{Introduction}
Quantum gravity appears to be experimentally inaccessible and will likely remain so for the foreseeable future. Viable theories must therefore satisfy certain theoretically motivated criteria. One of the most important of these is that a successful theory of quantum gravity must resolve the singularities that are inevitable in the classical theory. It must also describe the endpoint of black hole evaporation and thus resolve the information loss paradox. The CGHS\cite{cghs}  model was proposed in the early 90's as a theoretical laboratory for looking at precisely these issues. However, it did not resolve the singularity, nor was it solvable when the back reaction term was added. Shortly thereafter, Rousso and collaborators (RST)\cite{rst} modified the model by adding a local term to the anomaly, rendering it solvable. The singularity remained (see \cite{strominger} for an excellent early review.) .

The CGHS model was revived recently in an elegant paper by Ashtekar {\it et al} \cite{ashtekar1} who analyzed the quantum field theory version and used it to argue that information was not lost in the evaporation process. More recently, a numerical analysis of the semi-classical model \cite{ashtekar} obtained some interesting results by studying numerically the semi-classical dynamics of the CGHS model.

Clearly the issue of information loss is closely tied to singularity resolution, which requires the quantization of the gravitational sector of the theory.
Levanony and Ori (LO) \cite{LOa,LOb} have studied the CGHS model arguing that near the singularity the spacetime would be nearly homogeneous.  They examined the corresponding near singularity dynamics classically and quantum mechanically, verifying that the singularity was indeed resolved in the quantum version of their model.

In what follows, we re-examine the homogeneous black hole interiors of the radiating 2-D black hole along the lines of \cite{LOa, LOb} but with some modifications and extensions. Specifically, we look at the RST model as opposed to the CGHS model. As mentioned above, the extra local anomaly contribution renders the classical theory completely solvable and allows the construction of a much simplified  Hamiltonian.
At first, the RST model was cited as evidence that black hole information loss is possible in a self-consistent quantum theory. This is no longer generally believed to be the case.
As highlighted in \cite{strominger}, this model in its pure form suffers from an apparently serious shortcoming: it violates energy conservation in the form of an energy "thunderbolt" that emanates from the endpoint of the collapse/radiation process. This suggests, among other things, that the theory as given is not complete.  The incompleteness is also signified by the presence of the curvature singularity at intermediate stages of the evolution. It is therefore reasonable to see whether quantizing the gravitational degrees of freedom in the model can resolve the singularity and, ultimately, the other pathologies of the theory.

  The present paper is a first step towards a complete analysis of the quantum dynamics of the untruncated homogeneous theory. We first present the Lagrangian formulation of the homogeneous RST model . In order to work with only local covariant expressions we follow \cite{hayward95} and introduce an auxiliary field into the action. We then write down explicitly the full set of solutions and analyze the singularity structure. Next we isolate the black hole sector of interest and show that it consists of the static quantum black holes of \cite{birnir}, as expected by our requirement of homegeneity in the interior. These are radiating black holes in thermal equilibrium with incident radiation.
  We then focus on the dynamics of the dilaton field and show that it can be exactly generated by a Hamiltonian containing a time dependent potential and non-canonical kinetic energy terms. Next we study the near singularity dynamics of the dilaton, which turn out to be quantitatively different from its counterpart in the CGHS model. Finally we construct the Schr\"odinger quantum mechanics of the near singularity region using two methods, paying special attention to different boundary conditions and resulting self-adjoint extensions. We show that in both cases there is a range of extension parameters for which the Hamiltonian admits a bound state that could play the role of a stable black hole remnant.

The paper is organized as follows:
Section 2 presents the Lagrangian, field equations and exact solutions, including a discusion of the black hole sector of interest. The equivalent classical Hamiltonian analysis is relegated to an Appendix. Section 3 derives the effective Hamiltonian describing the dynamics of the dilaton and then presents the quantization of the near singularity theory via two different methods. Finally Section 4 closes with conclusions and prospects for future work.

\section{The Semi-Classical Theory}

\subsection{Action and Field Equations}

The theory starts with the classical CGHS action, which contains a dilaton $\phi$, a 2-D metric $g_{\mu\nu}$ and $N$ conformally coupled massless scalar fields \cite{cghs}:
\be
I_C[g_{ij},\phi]=\frac{1}{2\pi}\int d^2x\sqrt{-g}\left\{ e^{-2\phi}\left(R(g)+4\left[|\nabla\phi|^2+\lambda^2\right]\right)+\sum_{i=1}^N|\nabla f_i|^2\right\}.
\label{classical action}
\ee
Quantizing the scalar fields yields the well known conformal anomaly. In the large $N$ limit the one loop expression for the conformal anomaly is exact and contains the back-reaction on the metric of the quantized scalar fields. The semi-classical action we wish to consider describes the dynamics of the classical metric and geometry including the conformal anomaly terms. We therefore start with Hayward's local form of the RST action \cite{hayward95}, namely:
\bea
I[g_{ij},\phi,z]&=& I_C+I_A\none
   &=&\frac{1}{2\pi}\int d^2x\sqrt{-g}\left\{R(g)(e^{-2\phi}-\frac{\kappa}{2}\phi +\frac{\kappa}{2}z)\right.\none
 & &\left.+4\left[|\nabla\phi|^2+\lambda^2\right]e^{-2\phi}-\frac{\kappa}{4}|\nabla z|^2 +\sum_{i=1}^N|\nabla f_i|^2\right\}.
\label{action1}
\eea
where $\kappa\equiv N/12$ and we have set $\hbar=1$. The auxially field gives rise to the usual non-local anomaly term that corresponds to radiation back-reaction, while the second term in the coefficient of the Ricci scalar is the local term added by RST to make the theory solvable.

The equations of motion are given in Eqs. (3)-(5) of Hayward \cite{hayward95}.  With the $N$ sources $f_i(x)$ set to zero are:
\bea
&A^- R_{\mu\nu}+2 A^+\nabla_\mu\nabla_\nu\phi
-\frac{\kappa e^{2\phi}}{4}\left(2\nabla_\mu\nabla_\nu z+\nabla_\mu z\nabla_\nu z-\half g_{\mu\nu}|\nabla z|^2\right)=0;\label{met}\\
&A^+R+4\left(\Box\phi-|\nabla\phi|^2+\lambda^2\right)=0;\label{dil}\\
&\Box z+R=0.\label{z}
\eea
In the above
\be
A^\pm:=1\pm\frac{\kappa e^{2\phi}}{4}.
\ee
One can formally recover the usual non-local form of the action by writing the solution to (\ref{z}) as
\be
z = -\frac{1}{\Box R}
\ee
where $1/\Box$ refers to the scalar Green's function
and substituting this back into the $z\bx z$ term in the action (\ref{action1}), paying due attention to the boundary conditions. This heuristic argument is confirmed by a more careful analysis of the equations of motion(\cite{hayward95}). We note also that the second term in the coefficient of the Ricci scalar in the action (\ref{action1}) corresponds to the local term added by RST to the conformal anomaly. As we will see this term affects both the solvability of the theory (it was inserted for this purpose) and near singularity behaviour.

\subsection{Space of solutions}
We assume that the metric, dilaton $\phi$ and auxilliary field $z$ are functions of time only. In conformal gauge the metric is:
\be
ds^2=e^{2\rho(t)}(-dt^2+dx^2).
\label{eq:confy metric}
\ee
The nontrivial equations of motion are as follows.
The metric equations \req{met} reduce to
\bea
-A^-\ddot\rho+2A^+(\ddot\phi-\dot\phi\dot\rho)+\frac{\kappa e^{2\phi}}{2}(-\ddot z+\dot z\dot\rho-\frac{1}{4}\dot z^2 )=0;\label{mettt}\\
A^-\ddot\rho-2A^+\dot\phi\dot\rho+\frac{\kappa e^{2\phi}}{2}(\dot z\dot\rho-\frac{1}{4}\dot z^2 )=0.\label{metxx}\\
\eea
The dilaton equation of motion \req{dil} is
\be
A^+\ddot\rho-2\ddot\phi+2\dot\phi^2+2 \lambda^2 e^{2\rho}=0.\label{dilh}
\ee
Finally the $z$ equation of motion \req{z} is simply $\ddot z=2\ddot\rho$, with the general solution
\be
z(t)=2\rho(t)+z_1 t+z_0,\label{zsol}
\ee
where $z_1,z_0$ are integration constants.

If we substitute the latter into the trace of \req{met} (that is, into the difference between \req{mettt} and \req{metxx}), we find that $\ddot\phi=\ddot\rho$ and hence
\be
\rho(t)=\phi(t)+p_1 t +p_0,\label{rhosol}
\ee
with $p_1,p_0$ being integration constants.  Using the last two of the remaining independent equations of motion- that is, in one of \req{mettt} or \req{metxx}, and \req{dilh}, we find two second order nonlinear ordinary differential equations in $\phi(t)$:
\bea
A^-\ddot{\phi} - 2\dot{\phi}^2 -2 p_1 A^- \dot{\phi} + \frac{\kappa e^{2\phi}}{2}\left(p_1^2-\frac{z_1^2}{4}\right)=0
;\label{eq1}\\
-A^-\ddot\phi+2\dot\phi^2+2\lambda^2 e^{2(\phi+p_1 t +p_0)}=0\label{eq2}.
\eea

We note that the consistency condition for \req{eq1} and \req{eq2} is the first order nonlinear differential equation (i.e. constraint):
\be
-2p_1 A^-\dot\phi+e^{2\phi}\left[\frac{\kappa}{2}\left(p_1^2-\frac{z_1^2}{4}\right))+2\lambda^2 e^{2(p_1t+p_0)}\right]=0.\label{consisteq}
\ee
The general solution of \req{eq2} is
\be
\phi(t)=\half W(f(t))-\frac{1}{\kappa}\theta(t),\label{phisol}
\ee
where $W(x)$ is the LambertW function \cite{DLMF} defined implicitly by:
\be
W(x)e^{W(x)}=x,
\label{lambertw}
\ee
and
\bea
f(t)&:=&-\frac{4}{\kappa} e^{2\theta(t)/\kappa};\\
\theta(t)&:=&-2e^{-2\phi}-\kappa\phi\none
   &=&\frac{2\lambda^2}{p_1^2}e^{2(p_1 t+p_0)}+c_1 t+ \theta_0;\label{gensol}
\eea
The Lambert W function has a branch point singularity at $x=-1/e$. On the principal branch of $W(x)$ for which $W(x)>W(-1/e)$, $W(x)\to \infty$ as $x\to\infty$. On the other branch $W(x)\to -\infty$ as $x\to 0$.

In order that the consistency condition \req{consisteq} is satisfied, we need only
\be
p_1 c_1=\kappa\left(p_1^2-\frac{z_1^2}{4}\right).\label{consistsol}
\ee
Note that either the integration constant $z_0$ or $p_0$ can be eliminated by a trivial shift in the conformal time. We choose to set $p_0=0$.  Thus there are four remaining constants of integration $p_1,z_1,z_0,\theta_0$ remaining, which is consistent with the four dimensional reduced phase space presented in the Appendix.

There is another family of solutions, obtained by setting $p_1=0$ in \req{eq1}, \req{eq2} and \req{consisteq}.  These solutions have a curvature singularity, which, unlike the black hole solutions discussed in the next section, are not screened by a Killing horizon.

\subsection{Black Hole Sector}

We investigate whether or not the homogeneous geometry we have been analyzing can be the interior region of black hole. We find that this is the case precisely if the integration constants $p_1\neq0,z_1$ satisfy
\be
z_1=\pm2 p_1,\label{bhs}
\ee
that is, if the coefficient of the term linear in $t$ in the solution for $\theta(t)$ is zero. Moreover, the physical quantities associated with the black hole, i.e. the mass, horizon area, etc. are determined by the parameters $\theta_0$ and $z_1$ (or $p_1$) in the solution (\ref{gensol}). Below we summarize the reasons for this.

A sufficient condition for a black hole to exist is that there is a Killing horizon at some (perhaps infinite) value of the time coordinate $t=t_H$, such that the proper time for any observer to elapse from $t_H$ to the singularity (or any point earlier than the singularity) is finite, and such that the scalar curvature at $t_H$ is finite.

A necessary condition for a Killing horizon to exist is that $e^{2\rho(t)}\to 0^+$ as $t\to t_H$.  Using Eqs.(\ref{rhosol}), (\ref{phisol}) and (\ref{lambertw}) one can show that
\be
e^{2\rho(t)}=-\frac{4}{\kappa}\frac{e^{2(p_1t+p_0)}}{W(f(t))}.
\label{eq:rho1}
\ee
If $p_1>0$, then to get $e^{2\rho(t)}\to0$, we let $t\to-\infty$.  That is, $t_H=-\infty$.  In this limit, as long as $c_1>0$ it follows that $\theta(t)\to -\infty$, and hence that $f(t)\to 0^-$.  Thus $e^{2\rho(t)}$ goes to the indeterminate form $0/0$.  We change variables to $s:=e^{p_1 t}$ and write $C:=\kappa e^{2p_0}/4$.  Hence after using $W(x)e^{W(x)}=x$, and using L'Hospital's Rule, we find that
\be
e^{2\rho(t)}\to\frac{2 C}{a f_0} s^{2-a},
\ee
as $s\to 0^+$, where in addition $a:=2 c_1/p_1, f_0:=4 e^{2\theta_0}/\kappa$.  Hence $e^{2\rho(t)}$ converges to zero as $t\to-\infty$ if and only if $a\leq 2$, that is $\c_1\leq p_1$.  If we want this to converge for all $p_1>0$, then we need $c_1=0$, which as shown below, also ensures that the curvature at the horizon is finite. In this case, it can be seen from (\ref{gensol}) that $\theta \to \theta_0$ as $t\to \infty$. For future reference we define $f_0 \equiv f(\theta_0)$ and note that the limiting value of the Lambert W function at the horizon is $W(f_0)$.

The on-shell curvature scalar is:
\bea
R(t)&=&\frac{W(f(t))}{1+W(f(t))}\left[4\lambda^2
-\kappa^{-1}\frac{W(f(t))}{(1+W(f(t)))^2}e^{-2(p_1 t+p_0)} \left(c_1+(\frac{4\lambda^2}{p_1}e^{2(p_1 t +p_0)}\right)^2\right].
\eea
which indeed is finite as $t\to-\infty$ if $c_1=0$. In particular,
\be
R\to \frac{4}{\lambda^2} \frac{W(f_0))}{1+W(f_0)}.
\ee

Finally, we note that the proper time
\be
\tau(t_0)=\int_{-\infty}^{t_0}dt e^{2\rho(t)},
\ee
from the horizon to any point on the interior of the horizon, including the singularity, is also finite as long as $c_1=0$.

Ultimately, we would like to understand the semiclassical and quantum mechanical properties of these black holes.  The starting point for this is the determination of the mass and temperature of the black holes.  The most straightforward path to this consists of considering the theory in the region {\it exterior} to the horizon.  Hence we consider a metric ansatz of the form
\be
ds^2=e^{2\rho(x)}(-dt^2+dx^2).
\ee
We find that the {\it black hole} solution is of the same form, with $t$ replaced by $x$, and most importantly, with a minus sign in the term in the exponential of $x$, that is
\be
C(x):=e^{2\rho(x)}=-\frac{4}{\kappa}\frac{e^{2(-px+p_0)}}{W(f(x))},
\ee
where
\bea
f(x):&=&-\frac{4}{\kappa}e^{2\theta(x)/\kappa};\none
\theta(x):&=&-\frac{2\lambda^2}{p^2}e^{2(-px+p_0)}+\theta_0.
\eea
In terms of the corresponding integration constant in the interior homogeneous region, $p:=-p_1>0$.  With this choice of the sign of $p$, the event horizon occurs at the limit $x\to+\infty$, so that $C(x)\to 0^+$.  We also find that in this limit the curvature scalar is negative, and in the limit as $x\to-\infty$ -- the asymptotic region -- the curvature goes to zero.

We now transform to Schwarzschild  coordinates $t,r$ in which the metric takes the form.
\be
ds^2=-F(r)dt^2+\frac{1}{F(r)}dr^2,
\label{eq:schwarz}
\ee
where
\be
\frac{dr}{dx}=\pm C(x).
\label{eq:dr dx}
\ee
and $F(r(x))=C(x)$
We choose the minus sign in (\ref{eq:dr dx}) so that the asymptotic region is $r\to + \infty $, which in turn guarantees the correct sign for the mass.
The above can be solved:
\be
r=r_0+\frac{\kappa e^{2(-px+p_0)}}{8p}+\frac{p\kappa^2}{32\lambda^2}\left(W(f(x))+\frac{1}{W(f(x))}\right).
\ee
We can set $r_0=0$ without loss of generality, since it merely changes the origin of the radial coordinate.  In the limit as $x\to+\infty$, we find that
\be
r=r_H:=\frac{p\kappa^2}{32\lambda^2}\left(W_H(\theta_0)+\frac{1}{W_H(\theta_0)}\right),
\ee
where
\be
W_H(\theta_0):=W\left(-\frac{4}{\kappa} e^{2\theta_0/\kappa}\right).
\ee

The temperature of the black hole is easily calculated from the form (\ref{eq:schwarz}) by going to Euclidean time $t=-it_E$, and enforcing the regularity at the horizon by requiring that the Euclidean time coordinate is periodic there.  The temperature is the period of the Euclidean time. Hence:
\be
T_{BH}= \frac{1}{4\pi} \frac{dF}{dr} = \frac{1}{4\pi} \frac{dC}{dx}\frac{dx}{dr} = \frac{p}{2\pi}.
\label{eq:temperature}
\ee

Finally in order to identify the mass and make contact with the analyses in earlier work, we relate our solution to RST form\cite{rst}, in which the conformal mode of the metric is equal to the dilaton ($e^{2\tilde{\rho}} = e^{2\phi}$) and:
\bea
\Omega&=&  -\frac{\theta}{2}:=-2e^{-2\phi}-\kappa\phi\none
 &=& -\frac{\lambda^2}{\sqrt{\kappa}}\tilde{x}_+\tilde{x}_- + P \sqrt{\kappa}\ln\left(-\lambda^2 \tilde{x}_+ \tilde{x}_-\right)
   +\frac{M}{\lambda\sqrt{\kappa}}.
\label{eq:rst omega}
\eea.

In the metric (\ref{eq:confy metric}),
define $x_\pm = t\pm x$ so that:
$$
ds^2=-e^{2\rho(x_+ + x_-)}dx_+dx_-.
$$
RST \cite{rst} exploit the extra conformal coordinate invariance to go to coordinates in which $\rho = \phi$. We can do this by noting that for our solution:
$$
\rho-\phi = p_1 t = \frac{p_1}{2}(x_+ + x_-).
$$
By defining
\be
\tilde{x}_\pm = \frac{\pm 1}{p_1}e^{p_1 x_\pm},
\ee
we find
\be
2\rho\to 2\tilde{\rho} = 2\rho - p_1 (x_+ + x_-) = 2\phi.
\ee
In terms of the new coordinates,
\be
\tilde{x}_+\tilde{x}_- = \frac{1}{p_1^2} e^{p_1(x_++x_-)}=\frac{1}{p_1^2}e^{2p_1t}.
\ee
Replacing $\tilde{x}_\pm$ by $x_\pm$ in (\ref{eq:rst omega}) we get:
\begin{eqnarray}
\Omega &=& -\frac{\lambda^2}{p_1^2\sqrt{k}}e^{2p_1t}+P\sqrt{\kappa} \ln{\frac{\lambda^2}{p_1^2}e^{2p_1t}} + \frac{M}{\lambda\sqrt{\kappa}}\nonumber\\
&=&-\frac{\lambda^2}{p_1^2\sqrt{k}}e^{2p_1t}+P\sqrt{\kappa} \left\{\ln{\frac{\lambda^2}{p_1^2}}+{2p_1t}\right\} + \frac{M}{\lambda\sqrt{\kappa}}.
\end{eqnarray}
Thus our black hole sector solutions correspond with the RST solution for $P=0$ and $M=\theta_0\lambda\sqrt{\kappa}$. As noted by Birnir and Giddings\cite{birnir}, these solutions correspond physically to quantum black holes in thermal equilibrium with their environment at a fixed temperature that is independent of the mass.  These solutions are therefore static (in the exterior region) and have a regular horizon, as desired.
Given that we have from the outset restricted to homogeneous slicings our present formalism only allows consideration of regular horizons that are static with no net loss or gain of energy: if the interior is homogeneous and the horizon is regular, then the exterior must be static so that the horizon cannot shrink or grow. Thus the background and black hole temperatures must be the same. This balance is manifested by Eq. (23) which ensures that the solution does not diverge as $t\to -\infty$, i.e. at the horizon.

\section{Near Singularity Quantum Theory}

\subsection{Dilaton Hamiltonian}
Following \cite{LOa}, we define the variable:
\be
\tR=e^{-2\phi},
\ee
Using the identities:
\be
(\nabla \phi)^2 = \frac{1}{4}\frac{(\nabla \tR)^2}{\tR^2},
\ee
and
\be
\bx{\phi}= \frac{1}{2}\left(\frac{(\nabla \tR)^2}{\tR^2}- \frac{\bx{\tR}}{\tR}\right).
\ee
 one obtains from the dilaton equation (\ref{dil}):
\be
\bx{\tR} = -\frac{\kappa}{4}\frac{(\nabla \tR)^2}{\tR(\tR-\kappa/4)}+4\lambda^2 \frac{\tR^2}{\tR-\kappa/4}.
\label{tR}
\ee
Now with metric $ds^2=e^{2\rho}(-dt^2+dx^2)$, assuming homogeneity:
\bea
\bx{\tR}&=& \frac{1}{\sqrt{-g}}\partial_\mu\left(\sqrt{-g}g^{\mu\nu}\partial_\nu\tR\right)
\none
&=& -e^{-2\rho}\ddot{\tR}.
\eea
similarly
\bea
|\nabla \tR|^2 &=& \frac{1}{\sqrt{-g}}e^{2\rho}(-e^{-2\rho})(\dot{\tR})^2\none
&=& (-e^{-2\rho})(\dot{\tR})^2
\eea
Putting this into (\ref{tR}) gives:
\bea
\ddot{\tR} &=& -\frac{\kappa}{4}\frac{(\dot{\tR})^2}{\tR(\tR-\kappa/4)}-4\lambda^2 \frac{\tR^2e^{2\rho}}{\tR-\kappa/4}\none
  &=& -\frac{\kappa}{4}\frac{(\dot{\tR})^2}{\tR(\tR-\kappa/4)}-4\lambda^2 \frac{e^{2\rho-2\phi}\tR}{\tR-\kappa/4}\none
  &=& -\frac{\kappa}{4}\frac{(\dot{\tR})^2}{\tR(\tR-\kappa/4)}-4\lambda^2 \frac{e^{2(p_1t+p_0)}\tR}{\tR-\kappa/4},
\label{tR3}
\eea
where we have used the solution (\ref{rhosol}) to get the last line.

A key point is that the addition of the local term to the conformal anomaly has decoupled $\rho-\phi$ from $\tR$, yielding the conformal gauge, homogeneous solution (\ref{rhosol}). One can very that \req{tR3} for $\tR$ that can be generated by the following time dependent Hamiltonian:
\be
H_R=\frac{\Pi_{\tR}^2}{2}\left(\frac{\tR}{\tR-\kappa/4}\right)^2+4\lambda^2 e^{2(p_1t+p_0)}\left(\tR-\frac{\kappa}{4}\ln\tR\right).
\label{R hamiltonian}
\ee

The above Hamiltonian for the dilaton is exact. We now focus on the dynamics near the singularity $\tR=\kappa/4$, where the dominant term in the Hamiltonian is
\be
H_R=\frac{\kappa^2}{32}\frac{\Pi_{\tR}^2}{(\tR-\kappa/4)^2}.
\ee
This generates a near  singularity solution of the form:
\be
\tR-\frac{\kappa}{4}\propto(t-t_0)^{1/2},
\label{near sing}
\ee
which is different from the near singularity behaviour in \cite{LOa,LOb}, where $\tR-\kappa/4\propto (t-t_0)^{2/3}$. One can verify by doing a series expansion of the exact solution given in Section 2 that (\ref{near sing}) is the correct leading behaviour of $\tR$.


After a trivial shift of $R$, absorbing the numerical factor into the left hand side and dropping the tildes, the near singularity Hamiltonian (\ref{near sing}) is:
\be
H= \frac{\Pi^2}{R^2}.
\ee
\subsection{Quantization: Method 1}
Quantization of this Hamiltonian is straightforward. For the purpose of illustrating qualitatively what happens, we consider the simplest method, which is to
transform to $y=\half R^2$. Consider functions that are $L^2$ normalizable on $(0,\infty)$ with measure $\int dy = \int  R dR$\footnote{This is the natural parametrization invariant measure for (\ref{near sing}) if one interprets it as the Hamiltonian for a one dimensional ``sigma model'' with configuration space metric $g^{11} = 1/R^2$. We are grateful to J. Zanelli for pointing this out.}. The quantum Hamiltonian is then
\be
\hat{H}_1 = - \frac{d^2}{dy^2}.
\ee
Recalling that $R-\kappa/4>0$, this is just the quantum hamiltonian for a free particle on the half line. This system is well known. As described in section 6.2 of \cite{bonneau}, there exists a one parameter family of self-adjoints extensions of the Hamiltonian. Here we summarize the key points since they may not be familiar to everyone. The self-adjoint extensions correspond to choosing the following boundary conditions at the origin:
\be
\psi(0) = \lambda\psi'(0).
\label{boundary conditions}
\ee
With these boundary conditions there is a continuum of scattering states (i.e. positive energy eigenstates):
\be
\psi_k(y)= C \left[(-1+ik\lambda)e^{-iky}+(1+ik\lambda)e^{iky}\right],
\label{scattering states}
\ee
with energy:
\be
E_k = k^2.
\label{scattering energies}
\ee
Recall that the Hamiltonian is in the form of a free particle with $2m=1$.

From the states (\ref{scattering states}) it is possible to construct Gaussian, semi-classical wave-packets, as done by LO, that initially correspond to black hole boundary conditions and watch them evolve towards the singularity. Specifically we can take as a solution to the time-dependent Schrodinger equation:
\be
\psi(y,t)=A\int^\infty_0 dk C(k)\left[(-1+ik\lambda)e^{-iky}+(1+ik\lambda)e^{iky}\right] e^{-k^2 t},
\label{quantum solution}
\ee
where
\be
C(k)= e^{-2\alpha k^2}e^{iky_0},
\ee
which for suitable choice of initial parameters will approximate a wave packet centered at $y_0$, with width $1/\alpha$ (in position space) and initial momentum $0$. The latter is a necessary condition for the solution to start peaked near a black hole horizon \cite{LOa,LOb}. Singularity resolution, is self-evident in this case, since the wall is a perfect reflector for all values of the extension parameter\cite{bonneau}. The interesting feature that emerges, one that was not touched upon in (\cite{LOa, LOb}), is that for each $\lambda < 0$ there exists a unique normalizable bound state with probability amplitude peaked close to the singularity:
\be
\psi_b = \sqrt{\frac{1}{2|\lambda|}}e^{-y/|\lambda|},
\ee
with energy $E_b= -1/\lambda^2$. Reinstating the numerical factor previously absorbed into $H$, the physical bound state energy:
\be
E_b = -\frac{\kappa^2}{32\lambda^2}
\label{eq:Eb1}
\ee.
This could in principle describe a black hole remnant, since it corresponds to a stable, stationary localized ground state of the system.
\subsection{Quantization: Method 2}
Another approach is to consider wave functions normalizable on half line with measure $\int dR$. We choose the standard symmetric factor ordering for the Hamiltonian\footnote{GK thanks J. Louko for an invaluable discussion concerning this quantization scheme.}:
\bea
\hat{H}_2&=& - \frac{d}{dR}\frac{1}{R^2}\frac{d}{dR}\nonumber\\
&=& -R^2 \frac{1}{R^2} \frac{d}{dR}\frac{1}{R^2}\frac{d}{dR},\nonumber\\
&=& - y^{2/3} \frac{d^2}{dy^2}
\eea
where $y= \left(3^{-\half}R\right)^3\geq0$. In terms of the new coordinate the Hamiltonian is clearly symmetric with respect to the transformed measure:
\bea
\langle\chi|\hat H_2\psi\rangle&=&\tilde N\int^\infty_0 dR \chi^* \hat{H}_2 \psi=-N\int^\infty_0 dy y^{-2/3} y^{2/3} \chi^* \frac{d^2\psi}{dy^2}\nonumber\\
&=& -N\int^\infty_0 dy \frac{d^2\chi^*}{dy^2}\psi +N\int^\infty_0 dy
 \left((\chi^*)'\psi-\chi^* \psi' \right){}',
\eea
providing the boundary terms that appear on integration by parts vanish. Note that $'$  denotes differentiation with respect to $y$. We have absorbed factors of powers of 3 into the normalization constant $N:=3^{-\half}\tilde N$.  Assuming that we restrict to wave packets that vanish sufficiently fast at $y=\infty$, there is again a boundary condition at the origin:
\be
\left.\left((\chi^*)'\psi-\chi^* \psi' \right)\right|_0=0,
\ee
which requires (\ref{boundary conditions})
for some real $\lambda$ as in the previous method.

We first look for scattering states, which are solutions to
\bea
\hat{H}\psi(y) &=& E\psi(y)\nonumber\\
&\rightarrow& \psi''+y^{-2/3}E\psi=0,
\label{eigenvalue equation}
\eea
with $E=k^2>0$ (recall that the Hamiltonian is that of a free particle with $2M=1$). The solutions are:
\be
\psi_k(y) = C_+\sqrt{y} J_{3/4}(3 k y^{2/3}/2)+C_-\sqrt{y} J_{-3/4}(3 k y^{2/3}/2),
\label{basis functions}
\ee
where $J_\nu(Z)$ is the Bessel function of the first kind (\cite{DLMF}). We first check the asymptotic behaviour. Noting that (\cite{DLMF}):
\be
J_{\pm\nu}(Z)\to_{Z\to\infty} Z^{-1/2}\cos(Z\mp\nu\pi/2-\pi/4),
\ee
we find that
\be
\psi_k(y)\to \sqrt{y}(3k y^{2/3})^{-1/2}\cos(3 k y^{2/3}\mp\nu\pi/2-\pi/4).
\ee
This implies that (supressing constants):
\bea
\langle\psi|\psi\rangle &\to& \int dy y^{-2/3} y^{1/3}\cos^2(3k y^{2/3}\mp\nu\pi/2-\pi/4)\nonumber\\
  &~& \int d(y^{2/3}) \cos^2(3k y^{2/3}\mp\nu\pi/2-\pi/4),\nonumber
\eea
which shows that while non-normalizable at infinity, the solutions behave  asymptotically like plane waves and can therefore be used as a basis from which physical wave packets can be constructed.


We now proceed to the form of the scattering states as dictated by these boundary conditions. Using (10.7.3) from (\cite{DLMF})we find that near the origin the basis functions (\ref{basis functions}) go to:
\be
\psi_k(y) \to_{y\to0} \frac{3k}{4\Gamma(7/4)}(C_+y+C_-).
\ee
The boundary condition (\ref{boundary conditions}) therefore requires:
\be
C_- = \lambda C_+,
\ee
so that for each self-adjoint extensions the basis functions for the scattering states take the form:
\be
\psi_k(y) = C_+\left(\sqrt{y} J_{3/4}(3 k y^{2/3}/2)+\lambda\sqrt{y} J_{-3/4}(3 k y^{2/3}/2)\right).
\label{basis functions 2}
\ee
One can again readily construct semi-classical/Gaussian states from these basis functions and verify that they will indeed bounce from the origin.

We complete this section by looking for bound states $E=-k^2<0$. Using Maple, we find
\be
\psi_k(y)=\sqrt{y}\left[C_1I_{3/4}\left(\frac{3 k}{2}y^{2/3}\right)+C_2K_{3/4}\left(\frac{3 k}{2}y^{2/3}\right)\right],
\label{bessgen}
\ee
where $I_{3/4},K_{3/4}$ are modified Bessel functions of order $3/4$.  The term in $\psi_k(y)$ with the function $I_{3/4}$ is not normalizable with respect to the inner product $\int_0^\infty y^{-2/3} \bar\psi(y)\psi(y)$;  but the term in $K_{3/4}$ is.  We then find that the boundary condition (\ref{boundary conditions}) is satisfied for only one real value of $k=k_b$ for negative $\lambda$ given by
\be
k_b^{3/2}:=\frac{\sqrt{2}\Gamma^2(3/4)}{\sqrt{3}\pi \lambda}.
\ee
The normalized bound state wave-function is then
\be
\psi_b(y)=N_b(\lambda)\sqrt{y} K_{3/4}\left(\frac{3 k_b}{2}y^{2/3}\right).
\ee
The physical energy is again proportional to $\kappa^2/32$:
\be
E_b = - \frac{\kappa^2 k_b^2}{32} = -\frac{\kappa^2}{32}\left(\frac{\sqrt{2}\Gamma^2(3/4)}{\sqrt{3}\pi |\lambda|}\right)^{4/3}.
\label{eq:Eb2}
\ee

\section{Conclusions}

We have studied the interior classical and near singularity quantum dynamics of RST black holes. Via both a Lagrangian and Hamiltonian analysis we showed that one can readily isolate the static black hole sector of the theory and that it does, as expected, corresponded with the radiating quantum black hole of (\cite{birnir}). Using this as the starting point we investigated the dynamics of the dilaton in terms of conformal time and derived an effective Hamiltonian in which the dilaton was decoupled from the other modes apart from a time dependent coefficient in the potential.  Finally, following Levanony and Ori(\cite{LOa}) and (\cite{LOb}) we quantized the near singularity behaviour of the dilaton, paying close attention to boundary conditions. The Hamiltonian allowed a one parameter family of self-adjoint extensions that resolved the singularity.\footnote{We also note the work of A. Mikovic and his collaborators in computing one and two loop corrections to the CGHS model.  They showed that at the two loop level, the singularity was resolved. \cite{mik}.}. Of particular interest was the fact that there was a range of extension parameters admitting the existence of a single bound state, which could in principle play the role of a black hole remnant. We expect the existence of such bound states to survive in the full quantum theory, for suitable choices of the extension parameter.

The above results provide strong motivation for considering the full quantum theory of the black hole interior in more detail. In a future publication we hope to address this in two ways: via the Wheeler-DeWitt quantization of the Hamiltonian presented in the Appendix, and also via reduced quantization of the dilaton mode starting from the Hamiltonian (\ref{R hamiltonian}).

Finally, in order to address further fundamental issues, such as the quantum behaviour of radiating black holes not in equilibrium with their surroundings and the consistency of the full quantum theory, one needs to relax the assumption of homogeneity. This will be addressed in future work. \\[10pt]

\noindent
{\bf Acknowledgments:}\, We thank Abhay Ashtekar for suggesting the inclusion of the anamoly terms in the attempt to understand singularity resolution in quantum dilaton gravity. We are very grateful to Jorma Louko and Joey Medved for invaluable input. This research is supported in part by the Natural Sciences and Engineering Research Council of Canada. TT thanks the University of Manitoba for financial support. GK and TT thank the Centro de Estudios Cientificos, Chile for its kind hospitality during the completion of this work, as well as the members of the Institute for helpful comments and discussions.

\section{Appendix: Hamiltonian Analysis}
Here we consider only the homogeneous case, that is, all the metric components, and the fields $\phi,z$ depend on the local time coordinate $t$ only. One metric degree of freedom is used to set the lapse function, i.e. $g_{tx}$ to zero, so that
\be
ds^2=e^{2\rho(t)}(-\sigma^2(t)dt^2+dx^2).
\label{eq:metric}
\ee
We also normalize the action by dividing out the possibly infinite spatial volume.  Thus
\be
I[\beta, \phi,z]=\int dt\left(\Pi_\beta\dot\beta+\Pi_\phi\dot\phi+\Pi_z\dot z-H\right).
\ee
In the above, we have switched to a new configuration space variable $\beta:=\rho-\phi$ with conjugate momentum $\Pi_\beta$.  The Hamiltonian $H$ is
\be
H=\frac{2\pi\sigma}{B^2}\left[\kappa\left(\Pi_\phi+\frac{A}{\kappa}\Pi_z\right)^2+B\Pi_\beta(\Pi_\phi+2\Pi_z)-
\frac{\lambda^2}{\pi^2}B^2e^{2\beta}\right],\label{ham}
\ee
where the quantities $A,B$ are functions of $\phi$,
\bea
A:&=&4e^{-2\phi}+\kappa, \none
B:&=&4e^{-2\phi}-\kappa.
\eea
These quantities are related to the quantities $A^\pm$ in Section 3 by $A=4 e^{-2\phi}A^+,B=4 e^{-2\phi}A^-$.

Now we perform a canonical transformation which facilitates our eventual Dirac quantization of the model ${\phi,z}\to{y,w}$:
\bea
y&=&z+\frac{2e^{-2\phi}}{\kappa}-\phi;\qquad\Pi_y=-\frac{\kappa}{B(\phi)}(\Pi_\phi+2\Pi_z);\none
w&=&\kappa(z-2\phi);\qquad \Pi_w=\frac{(\Pi_\phi+\frac{A}{\kappa}\Pi_z)}{B(\phi)}.
\label{cant}
\eea

The Hamiltonian constraint has the simple form
\be
H=2\pi\sigma\left[\kappa\Pi_{w}^2-\frac{1}{\kappa}\Pi_\beta\Pi_y-\frac{\lambda^2}{\pi^2}e^{2\beta}\right].
\label{ham3}
\ee

We now solve the Hamiltonian equations of motion. We parameterize the fields with the proper time $T$ given by $dT=2\pi \sigma dt$.  We find, up to but not including imposition of the Hamiltonian constraint
\bea
w(T)&=&2\kappa P_w T+w_0;\qquad\Pi_w=P_w;\none
\beta(T)&=&-\frac{P_y}{\kappa}T+\beta_0;\qquad\Pi_\beta=-\frac{\lambda^2\kappa}{\pi^2 P_y}e^{-\frac{2P_y T}{\kappa}+2\beta_0}+P_\beta;\none
y(T)&=&-\frac{\kappa\lambda^2}{2\pi^2 P_y^2} e^{-\frac{2P_y T}{\kappa}+2\beta_0}-\frac{P_\beta}{\kappa}T+y_0;\qquad\Pi_y=P_y,
\label{eq:eom0}
\eea
where the six constants of integration are $w_0,P_w,y_0,P_y,\beta_0,P_\beta$.
The Hamiltonian constraint $H=0$ implies
\be
\kappa P_w^2-\frac{1}{\kappa}P_\beta P_y=0,\label{hamcon}
\ee
which can be used to eliminate one of the integration constants, say $P_\beta$ in terms of the other two.  Without loss of generality, the constant $\beta_0$ can be set to zero, since it is just a shift in the proper time.  Hence, there are four initial data parameters, $w_0,P_w,y_0,P_y$.

Note as well that the solutions above for $y(T)$ and $\Pi_\beta(T)$ are consistent with the only non-trivial dynamical equation that results from the Hamiltonian (\ref{ham3}), namely $\dot{y} = -\frac{2\pi \sigma}{\kappa}
\Pi_\beta$.

The parametrization can be transformed back to the original variables, but this transformation is singular when $B(\phi)=0$, i.e. at $\phi_s:=\half\ln{(4/\kappa)}$.  The range of $\phi$ is thus $\phi_s<\phi<\infty$.  The variables $\rho,z$ have the range from $-\infty$ to $0$, respectively $-\infty$ to $z_0$.

In terms of the original variables $\rho,\phi, z$, we find  that
$\theta(T)$ is given by the expression
\be
\theta(T):=(w_0-\kappa y_0) + \left(P_\beta +2\kappa P_{w}\right)T + \frac{\lambda^2 \kappa^2}{2\pi^2 P_y^2}e^{2\beta};
\label{theta}
\ee
\bea
\qquad\rho(T)&=&\beta+\phi \none
     &=& -\frac{P_y}{\kappa}T-\frac{1}{\kappa}\theta(T)+\half W\left(-\frac{4}{\kappa}e^{2\theta(T)/\kappa}\right);\none
\phi(T)&=&-\frac{1}{\kappa}\theta(T)+\half W\left(-\frac{4}{\kappa}e^{2\theta(T)/\kappa}\right);\none
z(T)&=& \frac{w}{\kappa} + 2\phi \none
   &=& \left(2y_0-\frac{w_0}{\kappa}\right)-2\left(\frac{P_\beta}{\kappa}+P_{w}\right)T
     -\frac{\lambda^2 \kappa}{\pi^2 P_y^2}e^{2\beta}+W\left(-\frac{4}{\kappa}e^{2\theta(T)/\kappa}\right).
\label{eq:eqsofmotion}
\eea
Notice that for this formulation the black hole sector is characterized by $P_\beta + 2 \kappa P_w = 0$.

The solutions \req{zsol}, \req{rhosol}, \req{phisol} and \req{gensol} of the covariant equations of motion can be reconciled with the solutions of the Hamiltonian equations of motion \req{eq:eqsofmotion} via the map $T = -\frac{t}{2\pi} $ and from the following identifications of integration constants:
\bea
\begin{array}{ll} 
p_0=\beta_0 \qquad&\kappa p_1=-P_y,\\
\theta_0=w_0-\kappa y_0\qquad&c_1=P_\beta+2\kappa P_w,\\
z_0=w_0-2\beta_0\qquad&\kappa z_1=2(\kappa P_w+P_y).
\end{array}
\eea
With these identifications, the consistency condition \req{consistsol} on the covariant equations of motion is equivalent to the Hamiltonian constraint \req{hamcon}.\\[10pt]

\end{document}